# An Approach to Avoid the Unreal High Flows on Congested Links and Investigates the Evolution of Congestion over Network

Sheng-Xue He[1*]

1. Business School, University of Shanghai for Science and Technology, Shanghai 200093, China;

**Abstract**: The unreal high flows may appear on the actually congested links in the result when a monotonically increasing link travel time function of flow volume is adopted in traffic assignment. The fixed link flow results of a static traffic assignment model (TAM) make it nearly impossible to investigate and make use of the actual evolution of congested zones over the network during the predetermined observation time period. Many methods, such as TAMs with side-constraints on link flow capacity and the pseudo dynamic traffic assignment based on day-by-day traffic, have been proposed to improve the reliability of the results of TAM, but cannot eliminate the problem. To resolve the above problems, we first uncover the origin of problem by analyzing the connection between the link travel time function and the fundamental diagram of traffic flow theory. According to the above connection a mapping is formulated to reflect the two branches of the fundamental diagram. Then with the assumption that the link flow states are given, new traffic assignment models are presented. To resolve the obtained non-convex user equilibrium model, a branch-and-bound algorithm is designed based on linearizing the nonlinear part of the objective function. At last, by repeatedly resolving the corresponding traffic assignment model with the renewed initial flow states of links, the evolution of congested zones during the observation time period can be reproduced and investigated. The numerical examples demonstrate the effectiveness of the new approach.

## 1. Introduction

The unreal high flows on the links that are congested in real life are commonly generated by solving all kinds of traffic assignment models. This is opposite to the fact when a link stays in a seriously congested flow state, the throughput flow during a given time period should be small. Except the above contradiction, the fixed link flows obtained by resolving static traffic assignment models cannot reflect the actual evolution of congested zones over the network. Observation tells us that the congested zones will dynamically expand and/or disperse during a given observation time period.

The unreal link flows make the urban transportation planning unreliable. Not only will the unreal flow result face serious challenges from all directions, but also the practicability of the other subsequent decisions is doubted. Since no existing theory supplies a satisfactory answer to the above confusion, practitioners and researchers feel baffling. The extreme response even asks whether transportation planning is science or not. Since the fixed flow pattern cannot reflect the changing of the real flow state over a network, many possible benefits from the observation of the above changing are lost, e.g. the adjustment of signal control schemes at intersections and more effective route guidance system.

[*] Tel.: +086-18019495040.
E-mail address: lovellhe@126.com

Many efforts have been made by researchers to deal with the above problems. Some researchers have been exploring the possibility of modifying the existing static traffic assignment models by adding side-constraints on the link flow capacities (Larsson and Patriksson, 1999; Larsson, et al. 2004; He, et al. 2011; Bliemer, et al. 2014; Tajtehranifard, et al. 2018). By limiting the maximal feasible flow on a link, extremely high link flows will be removed from the results. Although to resolve the modified model becomes more difficult, people are glad because they see the new method have the potential to remove the congestion from the network. But to our problem, nothing has changed. The unreal high link flow remains on the actually congested links in the result and the evolution of congested zones over the network during the observation period cannot be investigated by this new method.

Dynamic traffic assignment theory sounds a choice to cope with our problem (Zhong, et al. 2011; Lu, et al. 2015; Barthélemy and Carletti, 2017; Kostic, et al. 2017; Hoang, et al. 2018; Long, et al. 2018; Wang, et al. 2018; Zhu, et al. 2018; Zhao and Leclercq, 2018). Unfortunately, it is not. The reason lies in two aspects. On the one hand, to set up a practical DTA in practice, we need tons of data to calibrate the tons of coefficients. Except the appalling work, the result may be unconvincing. On the other hand, the fund limitation to a study of transportation planning always makes it nearly impractical to adopt such a method. Since a planning always needs to deal with the future situation, the data required for a DTA model in this situation are unavailable.

Recent years many researchers start to pay more attentions to the pseudo DTA based on day-by-day traffic data (Han and Du, 2012; He, et al. 2015; Guo, et al. 2016; Ma and Qian, 2017). Can this kind of DTA models supply us a satisfactory answer? The answer is still negative. By this theory, researchers wish to capture the path-choosing behaviors of travelers by investigating how they changing their route based on the past experiences. If their theoretical basis is similar to the existing static traffic assignment, these researches cannot naturally lead to a result where no unreal high link flow appears. The congested patterns can be compared on a daily basis. But it still fails to show the expanding and/or dispersing of congestion during a given observation time period.

In view of the above analyses, there is no way out of these practical and theoretical problems. Sounds all the known directions lead to a dead end. Until we turn our eyes to the origin of these problems, a rocky but feasible way shows up.

In this paper, we first try to uncover the origin of the problems. By analyzing the connection between the traffic flow-density-speed fundamental diagram and the travel time function of link flow, we note that the existing monotonically increasing travel time function can only correspond to the downside branch in the fundamental diagram. The congested upside branch is overlooked during the formulation of TAMs. This is the origin of our problem.

Assume that the links with congested flow are known, we can formulate new traffic assignment models with the monotonically decreasing travel time function as these links' new travel time function. Because the obtained system optimization model is convex, it can be resolved relatively easy by many methods. But the obtained user equilibrium model is non-convex. To resolve this non-convex model, a branch and band algorithm is proposed based on linearizing the nonlinear part of the objective function. Note that for a given time instant, the traffic flow state of a link is easy to be classified into uncongested or congested in practice.

To study evolution of congested zones with the given initial traffic flow states of links, we can repeatedly resolve the new formulated traffic assignment model with the renewed starting traffic flow states of links. The expanding and dispersing of congestion over the network can be reproduced in the above way.

The remaining part of this paper is organized as follows. Section 2 explains the reason why our problems arise by analyzing the connection between the commonly used link travel time function and the inverse lambda fundamental diagram of traffic flow theory. Section 3 presents the traffic assignment models with the given link flow states. Section 4 proposes a branch and bound algorithm to solve the non-convex user equilibrium model by linearizing the non-convex part of objective function. Section 5 presents a way to investigate the evolution of congested zones. Section 6 uses two road networks to test our new models and methods. Section 7 summarizes the main contents.

## 2. The Origin of Problem and the Inverse Lambda Fundamental Diagram

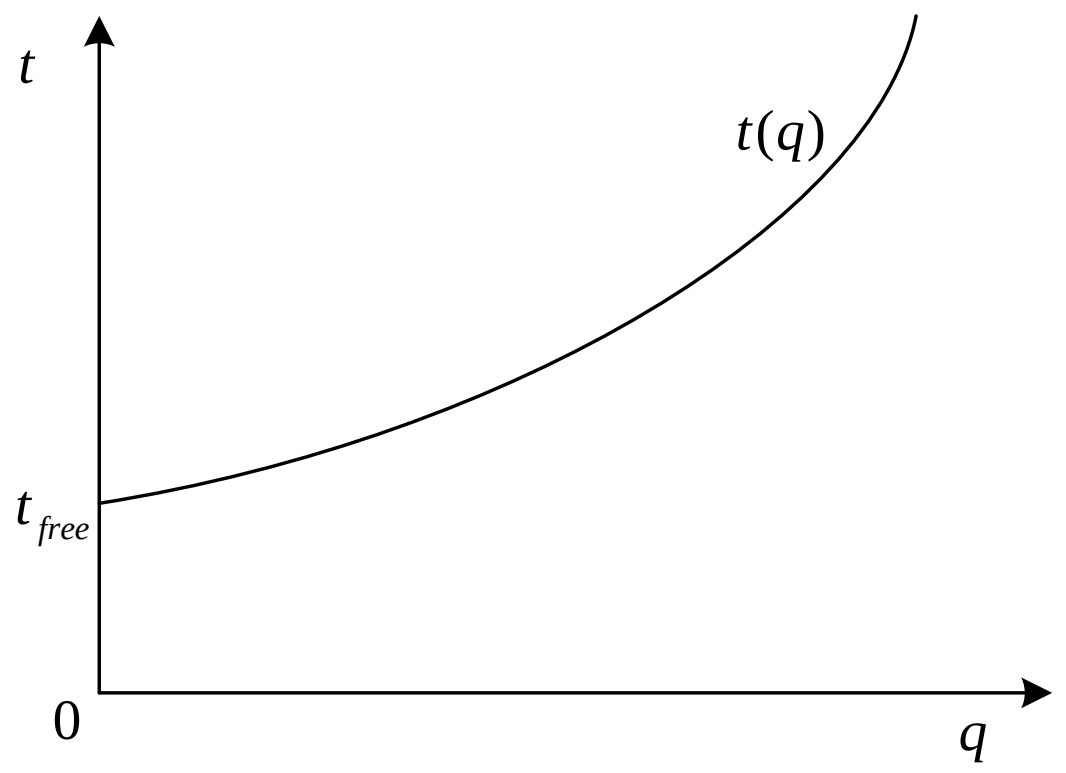

Fig. 1 A commonly used link travel time function of link flow

To simplify the subsequent expressions, we will use traffic flow or flow to stand for traffic flow rate. In some existing literatures, traffic flow rate is also called traffic volume. The measurement unit of traffic flow used in this paper is vehicles per hour (veh/hr). In Figure 1, a commonly used travel time function of link flow is plotted. With the increase of link flow $q$, the link travel time will gradually increase, too. When the traffic flow is zero, the corresponding travel time indicated by $t_{free}$ is called the travel time of free flow speed. At the first sight, this function seemingly embodies a correct relationship between link travel time and traffic flow. But this is not the fact. Only when the link remains in an uncongested state, the link travel time function will demonstrate the monotonically increasing feature. The above incorrect impression comes from the confusion of traffic flow with density. If we substitute density for flow, the impression that travel time will be lengthened with the increasing density conforms to the reality. But unfortunately, the travel time functions adopted by many existing studies are just like the one plotted in Figure 1. In our point of view, there are at least two shortages in these commonly used functions. On the one hand, these functions are lack of an upper limit to the feasible traffic flow. So the traffic assignment models based on such a link travel time function may produce some unreal large link flows. Some researchers have noticed this problem and then try to solve it by constructing new models with side constraints. Unavoidably, the resulted models are usually more complex and hard to be resolved effectively. On the other hand, these commonly used functions cannot properly reflect the relationship between travel time and link flow when a link stays in a congested state. According to modern traffic flow theory, when a link stays in a congested state, its travel time will increase with the decreasing traffic flow.

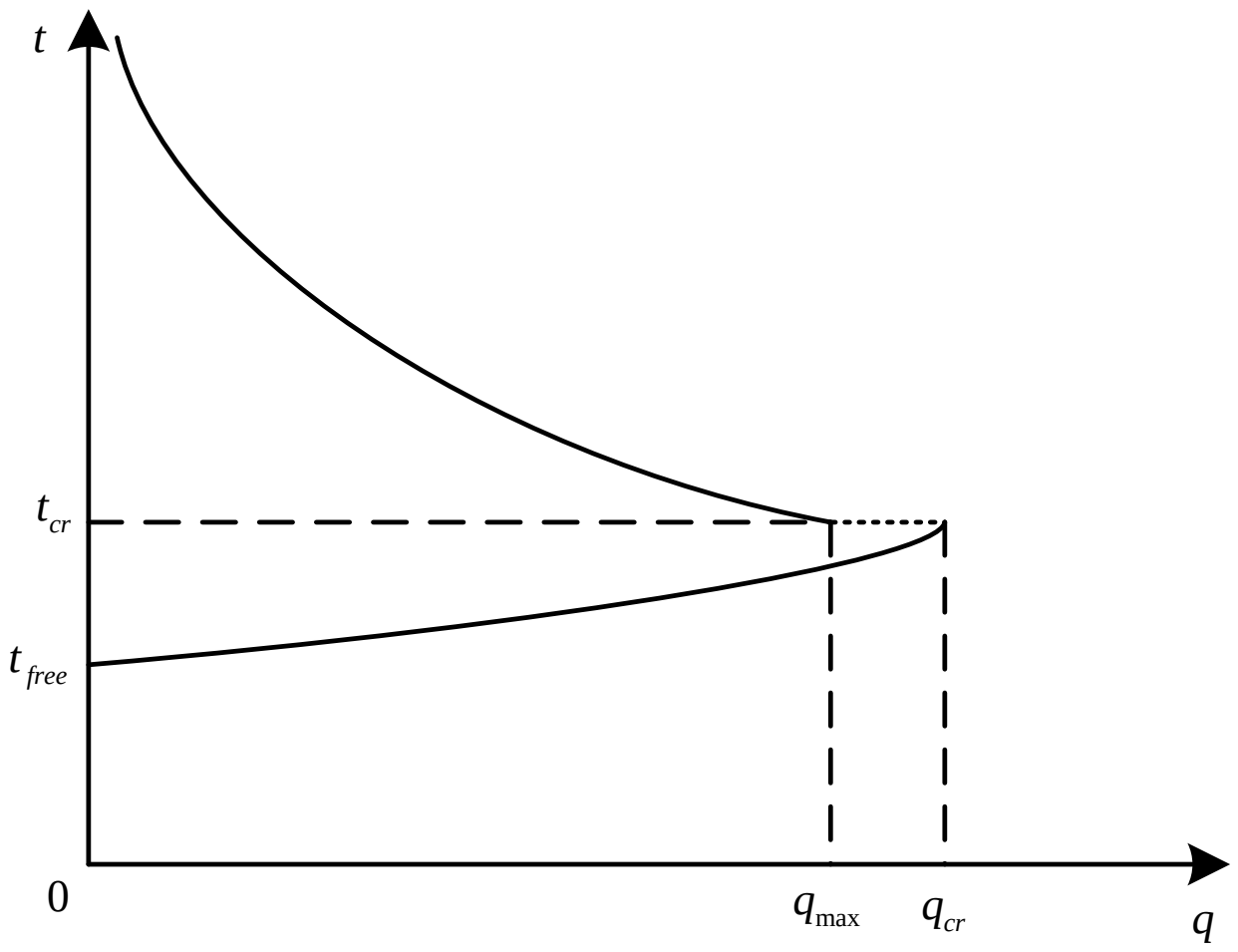

Fig. 2 The mapping form link flow to link travel time with two branches

In Figure 2, the mapping from link flow to link travel time with two branches is plotted. We assume that the biggest traffic flow for a link staying in the uncongested states is $q_{cr}$ defined as the critical flow. The biggest traffic flow for a link staying in the congested states is $q_{\max}$ defined as the maximum flow. The difference $q_{cr} - q_{\max}$ is generally called the capacity drop flow. For the mapping given in Figure 2, its downside branch corresponds to the uncongested states of traffic flow. In this branch, the travel time changes from $t_{free}$ to $t_{cr}$ and the corresponding traffic flow changes from 0 to $q_{cr}$. The upside branch is related to the congested states of traffic flow. In this branch, when the traffic flow changes from 0 to $q_{\max}$, the corresponding travel time changes from the positive infinite to $t_{cr}$. Different from the monotonically increasing feature of travel time function plotted in Fig. 1, the two branches of the new mapping have different changing trends. This mapping with two branches corresponds to the inverse lambda fundamental diagram.

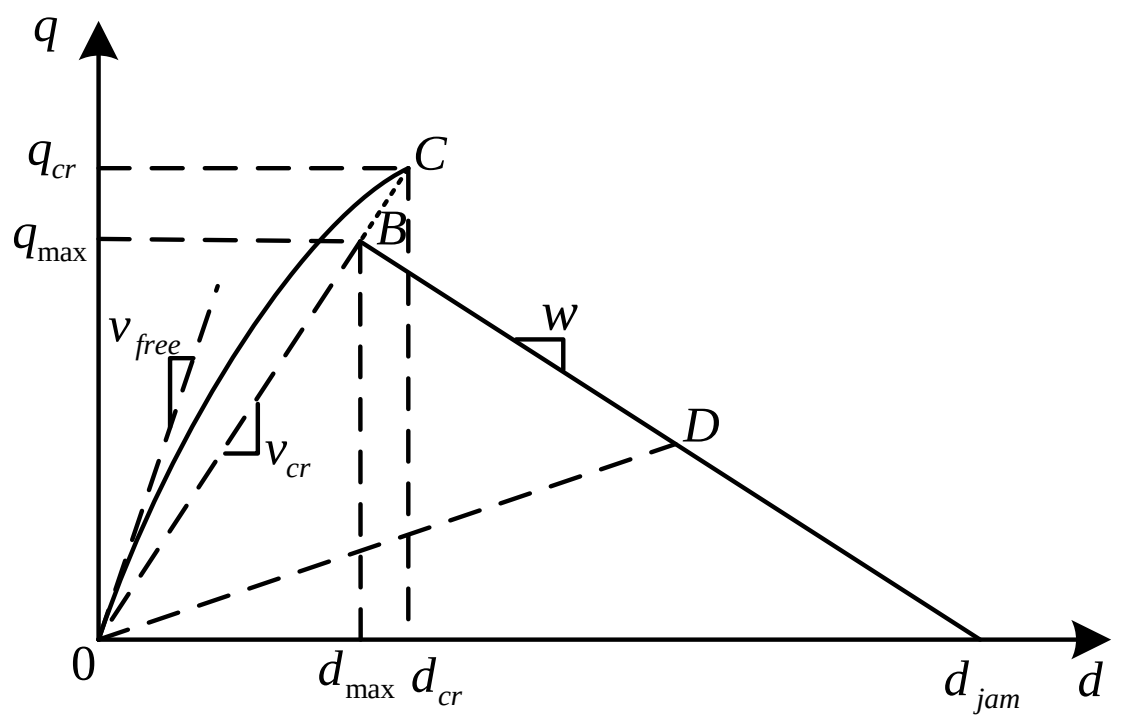

Fig. 3 The inverse lambda fundamental diagram

In Figure 3, a typical half curve fundamental diagram with inverse lambda shape is plotted. The solid curve connecting origin 0 to point $C$ is the uncongested branch corresponding to the uncongested states of traffic flow. The solid straight line passing points $B$ and $D$ is the congested branch corresponding to the congested states of traffic flow. This fundamental diagram embodies the capacity drop feature when the uncongested traffic flow state changes to congested. We assume that the vehicle speeds corresponding to the biggest uncongested flow $q_{cr}$ and the biggest congested flow $q_{\max}$ are the same as indicated by $v_{cr}$. It is well-known that the fundamental diagram of the modern three-phase traffic flow theory uses the specified areas to indicate the different traffic flow states. Different from the above diagram, the fundamental diagram with line or curve to indicate the traffic flow states can be more easily used to construct a travel time function. If the length of link in question is $l$, then we can obtain the following relations $t_{free} = l / v_{free}$ and $t_{cr} = l / v_{cr}$.

To simplify the subsequent analysis, we assume that the travel time function has the following form $t = t_{free} + \alpha q$ when the link stays at the uncongested state. But corresponding to the congested flow state, the form of travel time function becomes $t = \gamma + \beta q^{-1}$. Here the coefficients $\alpha$, $\beta$ and $\gamma$ are the undetermined constants for the link in question.

From the relations $t_{cr} = t_{free} + \alpha q_{cr} = l / v_{cr}$ and $t_{cr} = l / v_{cr}$, we can deduce the following equation $\alpha = (t_{cr} - t_{free}) / q_{cr} = l(1/v_{cr} - 1/v_{free}) / q_{cr}$. Since we assume that the line connecting points B and C in Figure 3 passes the original point, $v_{cr} = q_{\max} / d_{\max}$ holds. With above observation, the coefficient $\alpha$ can be determined by $\alpha = l(d_{\max} / q_{\max} - 1/v_{free}) / q_{cr}$. Conversely, we can use the given value of $\alpha$ to determine the curve corresponding to the connection between the traffic flow and traffic density in the uncongested states.

There is a correlation between the travel time function $t = \gamma + \beta q^{-1}$ and the linear part of inverse lambda fundamental diagram corresponding to the congested flow state. We can also obtain the following numerical relations $\gamma = l(d_{\max} - d_{jam}) / q_{\max}$ and $\beta = ld_{jam}$. The deducing process is as follows. Suppose that for any congested flow state $D$ as indicated in Figure 3, its corresponding flow and density are $q$ and $d$, respectively. We have $(d - d_{jam}) / q = (d_{\max} - d_{jam}) / q_{\max}$ based on the similar triangles. So we can further get $d / q = (d_{\max} - d_{jam}) / q_{\max} + d_{jam} / q$. Use the basic relations $t = l / v = ld / q$, we can obtain the travel time $t = l(d_{\max} - d_{jam}) / q_{\max} + ld_{jam} / q$ corresponding to the congested flow states. Comparing the above resulted travel time with the function $t = \gamma + \beta q^{-1}$, the coefficients $\beta$ and $\gamma$ can be determined.

**3. Modified Traffic Assignment Models**

The main notations about to be used in the modeling are introduced as follows.

- $a \in A$ is a directed link and $A$ is the set of all links.
- $n \in N$ is a typical node and $N$ is the set of all nodes.
- $A_n^+$ is the set of directed links which end at node $n$; in other words, if $a \in A_n^+$, node $n$ is the head node of link $a$.
- $A_n^-$ is the set of directed links which depart from node $n$; in other words, if $a \in A_n^-$, node $n$ is the tail node of link $a$.
- $b \in B$ is an Origin-Destination (OD) pair and $B$ is the set of all OD pairs.
- $n_b^+ \in N_B^+$ is the origin of OD pair $b$ and $N_B^+ \equiv \{n_b^+ \in N | b \in B\}$ is the set of all origins.
- $n_b^- \in N_B^-$ is the destination of OD pair $b$ and $N_B^- \equiv \{n_b^- \in N | b \in B\}$ is the set of all destinations.
- $q_b$ is the traffic demand between OD pair $b$.
- $x_a$ is the traffic flow on link $a$.
- $x_a^b$ is the part of traffic flow on link $a$ coming from the demand of OD pair $b$.
- $t_a^1(x_a)$ is travel time of link $a$ when the current traffic flow state is uncongested and the current flow is $x_a$ on link $a$.
- $t_a^0(x_a)$ is travel time of link $a$ when the current traffic flow state is congested and the current flow is $x_a$ on link $a$.

- $\delta_a \in \{0,1\}$ is the indicator of traffic flow state of link $a$. When the traffic flow state of link $a$ is uncongested, $\delta_a = 1$; or else, $\delta_a = 0$.
- $q_a^{\max}$ is the maximal flow when the traffic flow state is congested on link $a$.
- $q_a^{cr}$ is the critical flow or the biggest flow when the traffic flow state is uncongested on link $a$.
- $\Delta > 0$ is a given positive value used as the lower bound of traffic flow on a link when the traffic state of this link is congested.

The necessary constraints are as follows:

$$\sum_{a\in A_n^+} x_a^b - \sum_{a\in A_n^-} x_a^b = 0, \quad \forall n \in N / \{n_b^+, \mathrm{n}_b^-\}, b \in B, \tag{1}$$

$$\sum_{a\in A_n^+} x_a^b - \sum_{a\in A_n^-} x_a^b + q_b = 0, \quad \forall n = n_b^+, b \in B, \tag{2}$$

$$\sum_{a\in A_n^+} x_a^b - \sum_{a\in A_n^-} x_a^b - q_b = 0, \quad \forall n = n_b^-, b \in B, \tag{3}$$

$$\sum_{b\in B} x_a^b = x_a, \quad \forall a \in A, \tag{4}$$

$$x_a \le \delta_a q_a^{cr} + (1-\delta_a) q_a^{\max}, \quad \forall a \in A, \tag{5}$$

$$x_a \ge (1-\delta_a)\Delta, \quad \forall a \in A, \tag{6}$$

$$\delta_a \in \{0,1\}, \quad \forall a \in A. \tag{7}$$

Constraints (1), (2) and (3) are the traffic flow conservation equations corresponding to common node, origin and destination, respectively. Constraint (4) shows the relation between $x_a$ and $x_a^b$. In constraints (5) and (6), the upper and lower bounds of traffic flow $x_a$ are clarified. Constraint (7) limits the indicator $\delta_a$ to a 0-1 variable.

Corresponding to the well-known Wardrop's principles of travel route choice behaviors, two different objective functions can be constructed. In Eq. (8), the objective is related to the system optimization principle (the second Wardrop's principle). In Eq. (9), the objective is related to the user equilibrium principle (the first Wardrop's principle).

$$\min Z^{\mathrm{SO}} = \sum_{a\in A} x_a (\delta_a t_a^1(x_a) + (1-\delta_a) t_a^0(x_a)) \tag{8}$$

$$\min Z^{\mathrm{UE}} = \sum_{a\in A} \int_{(1-\delta_a)\Delta}^{x_a} (\delta_a t_a^1(w) + (1-\delta_a) t_a^0(w)) \mathrm{d}w \tag{9}$$

The System Optimization traffic assignment Model (SOM) consists of constraints (1~7) and objective function (8). The User Equilibrium traffic assignment Model (UEM) consists of constraints (1~7) and objective function (9).

**Proposition 1:** Assume that the travel time function of link $a \in A$ is given by $t_a = t_{free}^a + \alpha_a x_a$ when $\delta_a$ equals 1 and $t_a = \gamma_a + \beta_a x_a^{-1}$ when $\delta_a$ equals 0 and all the $\delta_a, \forall a \in A$ are given. In this case, the SOM is a convex quadratic programming.

**Proof:** For the given $\delta_a$, $\forall a \in A$, the set of constraints is a linear simplex. The part of objective function corresponding to the links with $\delta_a = 1$ is a convex quadratic function. The remaining part of objective function corresponding to the links with $\delta_a = 0$ is a linear function. With above observation, the statement of this proposition is proved.

**Proposition 2:** Assume that the travel time function of link $a \in A$ is given by $t_a = t_{free}^a + \alpha_a x_a$ when $\delta_a$ equals 1 and $t_a = \gamma_a + \beta_a x_a^{-1}$ when $\delta_a$ equals 0. All the $\delta_a, \forall a \in A$ are given and there is at least one link $c \in A$ such that $\delta_c = 0$. In this case, the UEM is a non-convex programming.

**Proof:** The adding term $\int_{\Delta}^{x_c} t_c^0(w))\mathrm{d}w = \int_{\Delta}^{x_c} (\gamma_c + \beta_c w^{-1})\mathrm{d}w$ with respect to $\delta_c = 0$ in the objective function is a concave function with respect to variable $x_c$. Due to this fact, the conclusion of this proposition can be deduced easily.

## 4. The Branch and Bound Algorithm for Non-convex UEM

From Proposition 1 we know SOM is a convex quadratic programming model. Now there are many effective and efficient methods of solving convex quadratic programming. Proposition 2 shows that UEM is a non-convex model with linear constraints. To such a model, we need to design a specified algorithm to find its optimal value and solution.

### 4.1. Linearizing the Non-convex Part of Objective Function

Assume that the set of links staying at congested state is denoted by $A_C \equiv \{a \in A | \delta_a = 0\}$. The number of elements in $A_C$ is denoted by $n_C$. From the objective function (9), we can see the non-convex part of UEM is $\sum_{a \in A_C} \beta_a \ln x_a$. In the congested flow state, the feasible range of link flow $x_a$ is the interval $[\Delta, q_a^{\max}]$. To simplify the subsequent expressions, we assign a unique serial number to every element of $A_C$. According to these serial numbers, use the corresponding link flows to construct a column vector $\hat{x} = (x_1, x_2, \cdots, x_{n_C})^{\mathrm{T}}$. Here the superscript "T" stands for the transposition operation. The convex part of objective function of UEM after removing the non-convex part is $Z^{\mathrm{UE}C} = Z^{\mathrm{UE}} - \sum_{a \in A_C} \beta_a \ln x_a$. We define sets $S \equiv \{\hat{x} \in R^{n_C} \,|\, l \le \hat{x} \le u\}$, $l = (l_1, \cdots, l_{n_C})^{\mathrm{T}} = (\Delta, \cdots, \Delta)^{\mathrm{T}}$, and $u = (u_1, \cdots, u_{n_C})^{\mathrm{T}} = (q_1^{\max}, \cdots, q_{n_C}^{\max})^{\mathrm{T}}$. Suppose that $S^k$ is any sub-box of $S$ and $S^k = \{\hat{x} \in R^{n_C} \,|\, l^k \le \hat{x} \le u^k\}$, where $l^k = (l_1^k, \cdots l_{n_C}^k)^{\mathrm{T}}$ and $u^k = (u_1^k, \cdots u_{n_C}^k)^{\mathrm{T}}$. It is easy to see that the convex hull of concave function $\beta_j \ln x_j$ in the interval $[l_j^k, u_j^k]$ is $\bar{y}_j = \frac{\beta_j (\ln u_j^k - \ln l_j^k)}{u_j^k - l_j^k}(x_j - l_j^k) + \beta_j \ln l_j^k$. This convex hull is a linear function and satisfies $\beta_j \ln x_j \ge \bar{y}_j$, $\forall x_j \in [l_j^k, u_j^k], j = 1, \cdots, n_C$.

If we substitute $\sum_{j=1}^{n_C} \bar{y}_j$ for the non-convex part $\sum_{a \in A_C} \beta_a \ln x_a = \sum_{j=1}^{n_C} \beta_j \ln x_j$ in the objective function of UEM, then the original non-convex programming (NCP) will be transformed to a convex quadratic programming (CQP). This CQP can supply a lower bound of the objective of the original model. Since the part of objective function related to uncongested states has been remained, the above transformation is partial linearization.

**4.2. Branch and Bound Algorithm Based on Sequential Convex Quadratic Programming**

The following branch and bound algorithm will determine the global optimal value of UEM by solving a series of CQPs and gradually improving the lower and upper bounds of the objective of UEM. Let $L_k$ denote the set of sub-boxes in which the global optimal solution may be at the $k^{\mathrm{th}}$ iteration. For any $\hat{S} \in L_k$, use $\mu(\hat{S})$ to denote the optimal value of CQP($\hat{S}$). Let $\mu_k = \min\{\mu(\hat{S}) \,|\, \hat{S} \in L_k\}$. $\mu_k$ will be a lower bound of UEM. If the optimal solution of CQP($\hat{S}$) is feasible with respect to the original UEM, then the upper bound $v_k$ of the objective value of UEM can be renewed. Assume that a sub-box $S^k$ satisfies $\mu(S^k) = \mu_k$. We can divide $S^k$ along its longest edge and obtain two relative small sub-boxes. We need to solve the CQPs in the new sub-boxes. The above process will continue until the terminal criterion is satisfied. Note that the variable $x_k$ to be used later stands for the vector consisting of link flow $x_a^b$. Obviously, the vector $\hat{x}$ can be obtained from $x_k$ and $\delta_a$.

The concrete process of the algorithm is given below.

**Step 1 Initialization**. Set the convergence criterion $\varepsilon > 0$ and let $k = 1$, $L_k = \{S\}$, $S^k = S$,

$\nu_k = \infty$. Solving the model CQP($S^k$), obtain its optimal solution $x_k$ and value $\mu(S^k)$. Let $\mu_k = \mu(S^k)$. If $x_k$ is feasible with respect to the original UEM, let $\nu_k = Z^{\mathrm{UE}}(x_k)$. If $\nu_k - \mu_k \le \varepsilon$, the algorithm stops. In this case, $x_k$ and $\nu_k$ are the optimal solution and value of UEM, respectively. Otherwise, continue to carry out the step 2.

**Step 2 Branching**. Along the longest edge of $S^k$, evenly divide $S^k$ into two part $S^{k_r}$ ($r = 1, 2$). Let $L_k := L_k - \{S^k\}$.

**Step 3 Bounding**. For any $r \in \{1, 2\}$, solve CQP($S^{k_r}$) and obtain the corresponding optimal solution $x_{k_r}$ and optimal value $\mu(S^{k_r})$. If $\mu(S^{k_r}) > \nu_k$, remove $S^{k_r}$; otherwise, let $L_k := L_k \bigcup \{S^{k_r}\}$. If $x_{k_r}$ is feasible for the original no-convex UEM, renew $\nu_k := \min\{\nu_k, Z^{\mathrm{UE}}(x_{k_r})\}$. And choose a $x_k$ such that $\nu_k = Z^{\mathrm{UE}}(x_k)$.

**Step 4 Termination check**. Let $L_{k+1} = L_k - \{\hat{S} \in L_k \,|\, \nu_k - \mu(\hat{S}) \le \varepsilon\}$. If $L_{k+1}$ is an empty set, the algorithm stops and the optimal solution and value of original UEM are $x_k$ and $\nu_k$, respectively; otherwise, let $k := k+1$ and $\mu_k = \min\{\mu(\hat{S}) \,|\, \hat{S} \in L_k\}$. Choose a sub-box $S^k$ such that $\mu(S^k) = \mu_k$ and return to Step 2.

**Proposition 3:** Assume that the globally optimal solution of the non-convex UEM exists. The branch and bound algorithm presented above will obtain this optimal solution in finite iterations or the limit of series $\{x_k\}$ as $k \to \infty$ will be the globally optimal solution.

**Proof:** If the algorithm terminates in finite iterations, the above conclusion is obvious. From the graph of function $\ln x$, we can see $\beta_j \ln x_j - \bar{y}_j \le \beta_j(\ln u_j^k - \ln l_j^k) \le \beta_j(u_j^k - l_j^k)$. Based on the feature of the convex hull of a concave function, there is the following relation $Z^{\mathrm{UE}}(S^k) \ge (Z^{\mathrm{UEC}}(S^k) + \sum_{j=1}^{n_C} \bar{y}_j)$ between the objective function of the original UEM and the one of the corresponding CQP. The following relation between the objective functions $Z^{\mathrm{UE}}(S^k) - (Z^{\mathrm{UEC}}(S^k) + \sum_{j=1}^{n_C} \bar{y}_j) = \sum_{j=1}^{n_C}(\beta_j \ln x_j - \bar{y}_j) \le n_C \max_j \beta_j(u_j^k - l_j^k)$ also holds. When infinite iterations are generated by the algorithm, in view of the feature of exhaustive enumeration of the box dichotomization, the limit $\lim_{k\to\infty}(u_j^k - l_j^k) = 0$ holds. To sum up the above analysis, we obtain $\lim_{x\to\infty}[Z^{\mathrm{UE}}(S^k)-(Z^{\mathrm{UEC}}(S^k)+\sum_{j=1}^{n_C}\bar{y}_j)]=0$ . The statement of this proposition is proved.

Though Proposition 3 has proved that the algorithm can obtain the globally optimal solution of UEM, the solution may not be unique in view of that the objective function of the relaxed CQP is not strict convex.

**5. The Way to Investigate the Evolution of Network traffic**

If the link flow states are given, the concrete network flow pattern satisfying one of the Wardrop's principles can be obtained by solving one of the corresponding models presented in section 3. But if the values of $\delta_a$ , $\forall a \in A$ are unknown, we cannot solve the mentioned model directly to determine the network flow pattern.

To carry out the traffic assignment, we generally assume that the traffic demands will keep steady during a relatively long time period. With this assumption, the link flows resulted from solving a traffic assignment model will reflect the steady network distribution of traffic flow with the given steady OD demands. During the evolution of network flow patterns, there must be a transition from uncongested to congested on some links when the OD demands are large enough and remain the same in a time period long enough. Sometimes we may observe a gradually extending zone consisting of congested links or a reverse process. In this section, we will focus on the way to reproduce the above transitions with the new models proposed in section 3. In the remainder of this section, the traffic assignment model to be used will not be specified. The choice depends on the aim of practical application.

**Definition 1**: Set $\delta_a=1$, $\forall a \in A$ and solve the traffic assignment model given in section 3. If the obtained link flows satisfy $x_a < q_a^{cr}, \forall a \in A$, we will say the network is a total uncongested traffic network. If there is at least one link $a \in A$ such that $x_a = q_a^{cr}$ , we will call the network a congested network with the first level congestion. If the model has no feasible solution, we call it a disabled network with the first level disability. For a congested network with the first level congestion, the link set $G^1 \equiv \{a \in A | x_a = q_a^{cr}\}$ will be called its first level bottleneck.

Let $\bar{G}^n \equiv G^1 \cup G^2 \cup \cdots \cup G^n$ and we will call $\bar{G}^n$ the $n$ th level congested zone of network. The link set $A/\bar{G}^n$ will be called the $n$ th level uncongested zone.

**Definition 2**: If we have obtained a network with the $n$ th level congestion, then we let $\delta_a=0$, $\forall a \in \bar{G}^n$ and $\delta_a=1$, $\forall a \in A/\bar{G}^n$ and solve the traffic assignment model presented in section 3. If the obtained link flows satisfy $x_a < q_a^{cr}$, $\forall a \in A / \bar{G}^n$, we will call the network a congested network with the final $n^{\text{th}}$ level congestion. If there is at least one link $a \in A / \bar{G}^n$ such that $x_a = q_a^{cr}$, the network will be called a congested network with the $(n+1)^{\text{th}}$ level congestion. If there is no feasible solution for the model, the network is a disabled network with the $(n+1)^{\text{th}}$ disability. For a congested network with the $(n+1)^{\text{th}}$ level congestion, the link set $G^{n+1} \equiv \{a \in A / \bar{G}^n \,|\, x_a = q_a^{cr}\}$ will be called its $(n+1)^{\text{th}}$ bottleneck.

In the above two definitions, the method of using the new traffic model to analyze the evolution of network flow pattern is implied. The method is summed up as follows.

**Step 1 Initialization**. Set $i=1$, $\bar{G}^0 = \varnothing$ and $\delta_a = 1$, $\forall a \in A$. Here $\varnothing$ is an empty set.

**Step 2 Solving traffic assignment model**. Solve the traffic assignment model (to choose UEM or SOM depends on the real application) with the given $\delta_a$.

**Step 3 Stopping check.** If there is no feasible solution to the model, the network is a disabled network with the $i^{\text{th}}$ disability. If the obtained optimal solution such that $x_a < q_a^{cr}$, $\forall a \in A / \bar{G}^{i-1}$, the network is a congested network with the final $(i-1)^{\text{th}}$ level congestion. Note that the congested network with the final $0^{\text{th}}$ level congestion is the total uncongested traffic network. If there is at least one line $\forall a \in A / \bar{G}^{i-1}$ such that $x_a = q_a^{cr}$, construct the sets of $G^i$ and $\bar{G}^i$ by $G^i \equiv \{a \in A / \bar{G}^{i-1} \,|\, x_a = q_a^{cr}\}$ and $\bar{G}^i \equiv \bar{G}^{i-1} \bigcup G^i$. If $G^i \neq \varnothing$, let $\delta_a = 1$, $\forall a \in \bar{G}^i$ and return to step 2; or else, stops.

The formation of the real traffic congested zone should parallelize the process of determining the different level congested zone described above. The real size of the congested zone will have a close connection to the length of the duration of OD demands. The process of dispersion of congestion will reverse the above forming process of congested zone. When there is no feasible solution to the traffic assignment model, it means the throughput capacity of the network is less than the OD demands. The traffic demands cannot be uploaded totally on the network and part of demands must queue at the origins waiting to be uploaded later.

## 6. Numerical Examples

Since the system optimization model is a convex quadratic programming, there are many solution algorithms which can solve them efficiently. So we will only focus on the non-convex user equilibrium model in this section.

In this section, we assume that the unit of time is hour (hr) and the unit of distance is kilometer

(km). The traffic flow is measured by vehicles per hour (veh/hr). The unit of speed is km/hr.

The branch and bound algorithm is realized by Lingo11. The sub-model of CQP is solved by the embedded program of Lingo11. So the efficiency of algorithm is closely related to the software used. We assign 0.001 to the stop criterion $\varepsilon$ and set $\Delta$=60 (veh/hr).

**6.1. The Way to Generate a Reasonable Link Travel Time Function**

To make the experimental traffic network about to be presented reasonable, we will create their parameters according to the fundamental diagram. Three steps will be followed. At first, we will use Lingo11 to randomly generate some parameters in their given ranges. These ranges should conform to the reality. Then we will deduce the other parameters with the known ones and the implicated relations among them. At last, the coefficients of travel time functions including $\alpha$, $\beta$ and $\gamma$ are determined by the equations given in Section 2.

To simplify the subsequent expressions, several notations need to be clarified firstly. $v_{cr}$ is the minimal speed when a link stays at uncongested state such that $v_{cr} = q_{cr} / d_{cr}$. Based on the assumption given in Section 2, we know that $v_{cr}$ is also the maximal speed when a link stays at congested flow state such that $v_{cr} = q_{\max} / d_{\max}$. $w$ is the backward shock wave speed. $r_{mc}$ is the rate of the difference of $q_{cr} - q_{\max}$ over $q_{\max}$. The value of $r_{mc}$ indicates the relative size of capacity drop with respect to the maximal flow $q_{\max}$.

The basic parameters are $v_{free}$, $v_{cr}$, $w$, $d_{jam}$ and $r_{mc}$. In this section, we assume that the feasible ranges of $v_{free}$, $v_{cr}$, $w$, $d_{jam}$ and $r_{mc}$ are intervals [60km/hr, 80km/hr], [40km/hr, 45km/hr], [15km/hr, 20km/hr], [110veh/km, 145veh/km] and [0.05, 0.08], respectively. Use the backward shock wave speed $w$ as an example to show how to randomly generate a feasible value for a basic parameter. Assume that $\varsigma$ is a number generated randomly from a uniform distribution over interval [0, 1]. Then we can obtain a concrete value of $w$ by $w$=15+(20-15)$\varsigma$. Here 15 and 20 are the lower and upper limits of the feasible range of $w$. The other basic parameters can be obtained in a similar way.

After the concrete values of basic parameters have been determined, we can compute the other parameters as follows. From the fundamental diagram in Figure 3, it is easy to get $w = q_{\max} / (d_{jam} - d_{\max})$ and $v_{cr} = q_{\max} / d_{\max}$. So the relation $w = d_{\max} v_{cr} / (d_{jam} - d_{\max})$ holds. Using this result, we can further obtain $d_{\max}$ by $d_{\max} = d_{jam} / (1 + v_{cr} / w)$. The relations including $q_{\max} = d_{\max} v_{cr}$, $q_{cr} = q_{\max}(1 + r_{cm})$ and $t_{free} = l / v_{free}$ are used to compute the values $q_{\max}$, $q_{cr}$ and $t_{free}$, respectively.

Based on the obtained values of parameters described above, the coefficients of link travel time function can be computed using the relations including $\alpha = l(d_{\max} / q_{\max} - 1/v_{free}) / q_{cr}$,

$\beta = ld_{jam}$ and $\gamma = l(d_{\max} - d_{jam}) / q_{\max}$.

**6.2. Test the Branch and Bound Algorithm**

To obtain the final result of the following example by our new algorithm, computational time of 27s is required. But if we solve the non-convex model directly using the embedded algorithm of Lingo11, the computational time is about 4s.

In Figure 4, a traffic network with 10 nodes is given. Assume that all the horizontal links are 2km. All the upper level vertical links are 1km and the lower level vertical links are 2km. The lengths of slant links have been labeled near to the corresponding links in the figure.

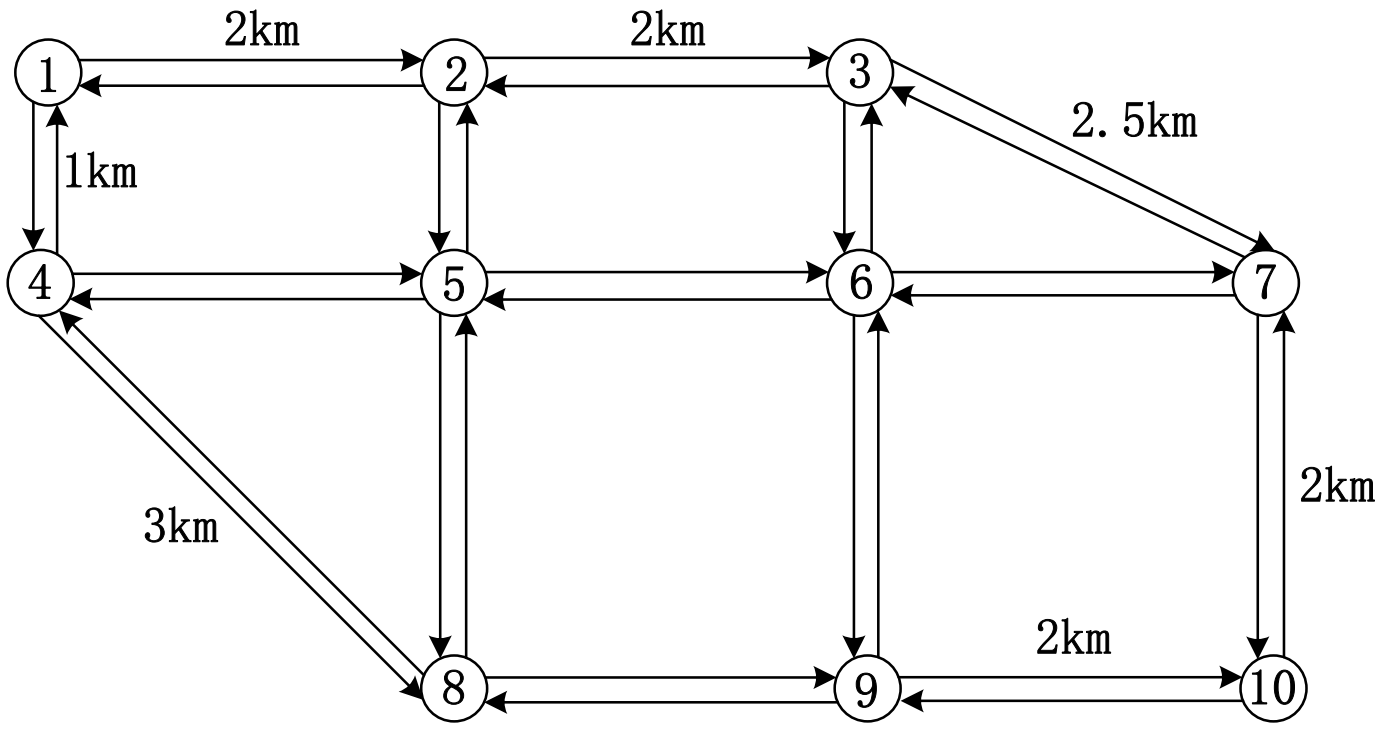

Fig. 4 Traffic network with 10 nodes

Table 1

The OD demands (veh/hr).

| Origin | Destination | | | | | | | | | |
|---|---|---|---|---|---|---|---|---|---|---|
| | 1 | 2 | 3 | 4 | 5 | 6 | 7 | 8 | 9 | 10 |
| 1 | 0 | 20 | 50 | 430 | 190 | 400 | 250 | 230 | 80 | 50 |
| 2 | 20 | 0 | 20 | 50 | 400 | 200 | 250 | 120 | 150 | 130 |
| 3 | 50 | 20 | 0 | 160 | 360 | 150 | 220 | 140 | 200 | 90 |
| 4 | 430 | 50 | 160 | 0 | 430 | 100 | 360 | 130 | 30 | 50 |
| 5 | 190 | 400 | 360 | 430 | 0 | 200 | 400 | 340 | 330 | 360 |
| 6 | 400 | 200 | 150 | 100 | 200 | 0 | 140 | 220 | 120 | 70 |
| 7 | 250 | 250 | 220 | 360 | 400 | 140 | 0 | 320 | 230 | 60 |
| 8 | 230 | 120 | 140 | 130 | 340 | 220 | 320 | 0 | 40 | 80 |
| 9 | 80 | 150 | 200 | 30 | 330 | 120 | 230 | 40 | 0 | 80 |
| 10 | 50 | 130 | 90 | 50 | 360 | 70 | 60 | 80 | 80 | 0 |

Table 2

Coefficients, parameters and flows of links of Figure 4.

| $a$ | $\gamma$ | $\beta$ | $\alpha$ | $q_{\max}$ | $q_{cr}$ | $\delta_a$ | $x_a(1)$ | $x_a(2)$ |
|---|---|---|---|---|---|---|---|---|
| (1,2) | -0.12651 | 273.228 | 1.26E-05 | 1567.173 | 1662.683 | 1 | 1170.536 | 1662.683 |
| (1,4) | -0.05112 | 127.8856 | 4.51E-06 | 1729.772 | 1854.16 | 0 | 1672.628 | 539.4359 |
| (2,1) | -0.12083 | 247.6413 | 1.47E-05 | 1453.106 | 1541.664 | 1 | 1228.836 | 747.915 |
| (2,3) | -0.10767 | 230.1844 | 1.02E-05 | 1504.091 | 1612.25 | 1 | 1612.25 | 1612.25 |
| (2,5) | -0.05771 | 141.3638 | 4.74E-06 | 1723.192 | 1848.118 | 0 | 904.1607 | 1723.192 |
| (3,2) | -0.10352 | 228.8147 | 1.11E-05 | 1520.592 | 1629.938 | 1 | 1629.938 | 1629.938 |
| (3,6) | -0.05943 | 119.842 | 6.82E-06 | 1452.813 | 1530.484 | 1 | 1530.484 | 1530.484 |
| (3,7) | -0.16299 | 361.0587 | 1.32E-05 | 1638.775 | 1759.955 | 1 | 1470 | 1640.658 |
| (4,1) | -0.06638 | 136.614 | 6.12E-06 | 1506.029 | 1614.327 | 1 | 1614.327 | 1454.204 |
| (4,5) | -0.11383 | 284.0972 | 1.1E-05 | 1779.058 | 1874.171 | 0 | 115.8393 | 753.677 |
| (4,8) | -0.15125 | 405.0306 | 1.52E-05 | 1802.991 | 1937.373 | 0 | 1774.546 | 163.64 |
| (5,2) | -0.05112 | 115.0922 | 4.5E-06 | 1551.536 | 1664.017 | 0 | 944.7734 | 790.7362 |
| (5,4) | -0.12059 | 282.7275 | 1.26E-05 | 1682.153 | 1772.085 | 1 | 1772.085 | 1772.085 |
| (5,6) | -0.1304 | 265.2706 | 1.23E-05 | 1479.256 | 1559.21 | 0 | 1303.204 | 1479.256 |
| (5,8) | -0.10476 | 223.5967 | 8E-06 | 1494.854 | 1592.042 | 0 | 60 | 1494.854 |
| (6,3) | -0.05771 | 128.5704 | 6.6E-06 | 1581.183 | 1665.718 | 1 | 1495.06 | 1665.718 |
| (6,5) | -0.12436 | 239.6839 | 1.03E-05 | 1414.717 | 1491.182 | 0 | 1289.42 | 1414.717 |
| (6,7) | -0.10083 | 222.227 | 8.89E-06 | 1509.761 | 1607.918 | 0 | 60 | 60 |
| (6,9) | -0.11863 | 274.7702 | 1.23E-05 | 1651.489 | 1776.971 | 1 | 1509.079 | 1100.849 |
| (7,3) | -0.15074 | 293.6677 | 1.46E-05 | 1418.239 | 1523.112 | 1 | 1523.112 | 1523.112 |
| (7,6) | -0.1061 | 266.6403 | 8.65E-06 | 1730.054 | 1842.534 | 0 | 179.3317 | 179.3317 |
| (7,10) | -0.12228 | 231.7266 | 1.19E-05 | 1382.823 | 1488.703 | 1 | 1067.557 | 1067.557 |
| (8,4) | -0.18654 | 392.8358 | 1.72E-05 | 1519.254 | 1632.487 | 0 | 60 | 60 |
| (8,5) | -0.10207 | 265.2706 | 8.31E-06 | 1770.643 | 1885.762 | 0 | 1770.643 | 1645.345 |
| (8,9) | -0.11696 | 276.1399 | 1.06E-05 | 1685.491 | 1814.546 | 1 | 1814.546 | 1638.494 |
| (9,6) | -0.10476 | 222.227 | 1.22E-05 | 1480.968 | 1593.494 | 1 | 1340.539 | 1052.212 |
| (9,8) | -0.11208 | 274.7702 | 1.02E-05 | 1731.468 | 1864.043 | 1 | 1810.643 | 1685.345 |
| (9,10) | -0.10067 | 233.0963 | 6.74E-06 | 1605.58 | 1697.956 | 1 | 1550 | 1379.342 |
| (10,7) | -0.11863 | 249.1835 | 1.29E-05 | 1492.835 | 1607.139 | 1 | 1240 | 1069.342 |
| (10,9) | -0.12811 | 231.7266 | 1.21E-05 | 1302.612 | 1377.557 | 1 | 1377.557 | 1377.557 |

Assume that any one node can be an origin and at the same time a destination. So the total number of valid OD pairs is 90. The OD demands are presented in Table 1. The total demand is 16869veh/hr. We also assume that the free flow speed $v_{free}$ is 80km/hr.

In Table 2, the related coefficients and parameters of links of the traffic network in Figure 4 and the computed link flows are presented. Links are indicated by a pair of ordered nodes which are the tail node and the head node, respectively. The column titled by $x_a(1)$ stores the link flows resulted from our new branch and bound algorithm. The column of $x_a(2)$ lists the link flows resulted from directly using the Lingo to solve the non-convex model.

Table 3

The upper and lower bounds of objective function.

| $k$ | 1 | 2 | 3 | 4 | 5 | 6 | 7 | 8 | 9 |
|---|---|---|---|---|---|---|---|---|---|
| $\mu_k$ | 16304.02 | 16644.18 | 16778.48 | 16778.48 | 17134.85 | 17134.85 | 17134.85 | 17134.85 | 17134.85 |
| $\nu_k$ | 17898.35 | 17398.1 | 17398.1 | 17398.1 | 17398.1 | 17398.1 | 17398.1 | 17398.1 | 17398.1 |

Table 4

Objectives of sequentially encountered CQPs and UEMs.

| No. | OCQP | OUE | No. | OCQP | OUE |
|---|---|---|---|---|---|
| 1 | 16304.02 | 17898.35 | 11 | 16830.5 | 17866.44 |
| 2 | 16430.07 | 17637.44 | 12 | 17225.23 | 18047.2 |
| 3 | 16644.18 | 17398.1 | 13 | 16910.8 | 17866.44 |
| 4 | 16676.4 | 17866.44 | 14 | 17008.92 | 17866.44 |
| 5 | 16778.48 | 17658.65 | 15 | 17048.08 | 17877.99 |
| 6 | 17133.97 | 18606.15 | 16 | 17048.08 | 17877.99 |
| 7 | 16676.4 | 17866.44 | 17 | 17790.41 | 18649.78 |
| 8 | 16720.46 | 17866.44 | 18 | 17790.41 | 18649.78 |
| 9 | 17134.85 | 18136.24 | 19 | 18226.45 | 19088.92 |
| 10 | 17247.06 | 18136.24 | | | |

In Table 3, the changing of the upper and lower bounds of objective with the increasing iterations during the process of computing $x_a(1)$ is summarized. The new algorithm terminates at the 9$^{th}$ iteration (19 times to solve the convex quadratic programming sub-model) after obtain the global optimal value 17398.1. The optimal objective value obtained by directly solving the non-convex model using Lingo is 17613.89. In Table 2, the corresponding link flows are presented in columns of $x_a(1)$ and $x_a(2)$, respectively. The objective values of sub-model (OCQP) related to the 19 times of solving sub-model and their corresponding objective values of original UE model (OUE) are summed up in Table 4. In view of the final objective value, the above results demonstrate that our new algorithm outperforms the embedded algorithm of Lingo 11.

### 6.3. Analyze Evolution of Network Flow Patterns

In Figure 5, a simple traffic network with 7 nodes is given. We assume that nodes 1 and 3 are origins and nodes 5, 6 and 7 are destinations. There are four feasible OD pairs including (1, 6), (1, 7), (3, 5) and (3, 7). The corresponding OD demands of above four OD pairs are 1200, 1800, 500 and 600 (veh/hr), respectively. The related coefficients and parameters of links are summed up in Table 5.

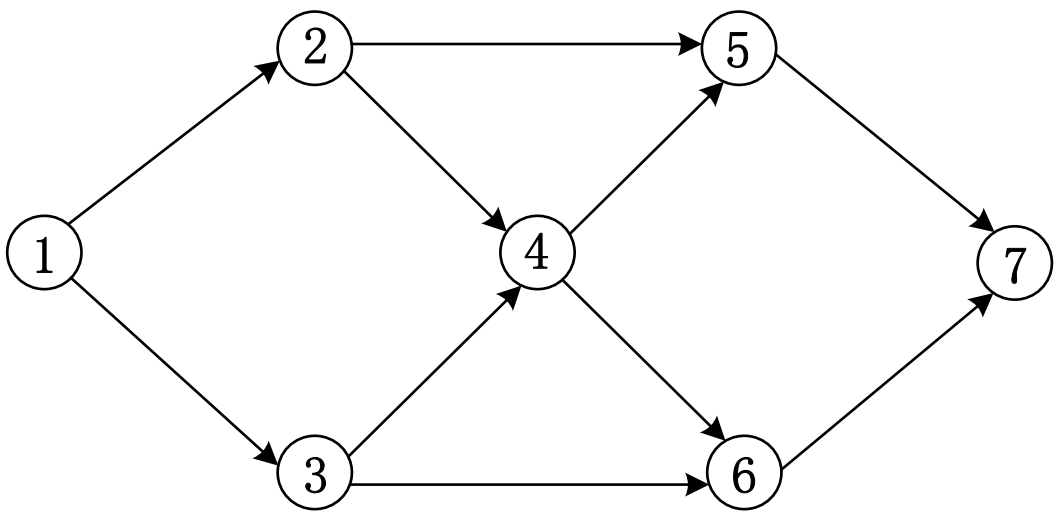

Fig. 5 Traffic network with 7 nodes

Table 5

Coefficients, parameters of links of Figure 5.

| $a$ | $\mathcal{L}$ | $\gamma$ | $\beta$ | $\alpha$ | $v_{free}$ | $q_{\max}$ | $q_{cr}$ |
|---|---|---|---|---|---|---|---|
| (1,2) | 3 | -0.18977 | 409.842 | 1.88E-05 | 74.2155 | 1567.173 | 1662.683 |
| (1,3) | 4 | -0.20704 | 504.782 | 2.65E-05 | 75.18124 | 1648.38 | 1733.151 |
| (2,5) | 2 | -0.11541 | 282.7275 | 9.48E-06 | 64.24015 | 1723.192 | 1848.118 |
| (2,4) | 2 | -0.10628 | 233.5645 | 9.57E-06 | 63.27441 | 1524.144 | 1603.421 |
| (3,4) | 2 | -0.11066 | 257.1408 | 9.94E-06 | 64.24015 | 1616.875 | 1734.094 |
| (3,6) | 3 | -0.17829 | 359.5259 | 2.05E-05 | 79.25247 | 1452.813 | 1530.484 |
| (4,5) | 3 | -0.17074 | 426.1458 | 1.65E-05 | 79.25247 | 1779.058 | 1874.171 |
| (4,6) | 2 | -0.12436 | 287.4773 | 1.12E-05 | 73.29905 | 1672.891 | 1796.594 |
| (5,7) | 3 | -0.15915 | 366.6505 | 1.36E-05 | 68.31138 | 1617.017 | 1737.538 |
| (6,7) | 4 | -0.20166 | 444.4541 | 1.78E-05 | 62.35796 | 1509.761 | 1607.918 |

Table 6

The equilibrium link flows under different link states.

| $a$ | 1 | | 2 | | 3 | | 4 | |
|---|---|---|---|---|---|---|---|---|
| | $\delta_a$ | $x_a$ | $\delta_a$ | $x_a$ | $\delta_a$ | $x_a$ | $\delta_a$ | $x_a$ |
| (1,2) | 1 | 1662.683 | 0 | 1567.173 | 0 | 1266.849 | 0 | 1567.173 |
| (1,3) | 1 | 1337.317 | 1 | 1432.827 | 1 | 1733.151 | 0 | 1432.827 |
| (2,5) | 1 | 1662.683 | 1 | 1567.173 | 1 | 1266.849 | 1 | 1567.173 |
| (2,4) | 1 | 0 | 1 | 0 | 1 | 0 | 1 | 0 |
| (3,4) | 1 | 906.8331 | 1 | 1734.094 | 0 | 1616.875 | 0 | 1616.875 |
| (3,6) | 1 | 1530.484 | 0 | 798.7333 | 0 | 1216.276 | 0 | 915.9518 |
| (4,5) | 1 | 500 | 1 | 647.604 | 1 | 784.75 | 1 | 625.3253 |
| (4,6) | 1 | 406.8331 | 1 | 1086.49 | 1 | 832.1252 | 1 | 991.5499 |
| (5,7) | 1 | 1662.683 | 1 | 1714.777 | 1 | 1551.599 | 1 | 1692.498 |
| (6,7) | 1 | 737.3169 | 1 | 685.223 | 1 | 848.4012 | 1 | 707.5017 |

According to the method of investigating the evolution of network flow patterns presented in section 5, we first assume that all the links stay at uncongested states. In this situation, we can obtain the equilibrium link flows listed in the column of $x_a$ under the first scenario in Table 6. Notice that the link flows on links (1, 2) and (3, 6) reach their upper limit (the critical flow) in uncongested states. So assume that these two links enter the congested flow states. In this situation, the obtained link flows are summed up in the column of $x_a$ under the second scenario in Table 6. We notice that the traffic flow on link (3, 4) reaches its uncongested upper limit (the critical flow) now. Based on this observation, we can assume that link (3, 4) enters its congested flow state so as to obtain the third scenario. The values of $\delta_a$ under this scenario are also summed up in Table 6. The computational result with respect to the third scenario shows that link (1, 3) reaches its congested flow state. Set the value of $\delta_a$ of link （1, 3）to 0, we get the fourth scenario. The resulted link flows under this scenario are stored in the column of $x_a$ under the 4$^{th}$ scenario. This time we find that no new link reaches its congested flow state. Based on this observation, we can say this network is a congested network with the final 3$^{th}$ congestion. The 3$^{th}$ level congested zone of this network consists of links (1, 2), (3, 6), (3, 4) and (1, 3). The objective values of UEM of above four scenarios are 544.22, 5418.69, 7131.43 and 10362.75, respectively.

## 7. Conclusions

The unreal high flows on the actually congested links come from some improper travel time functions used in the existing TAMs. This imperfection also lead to the powerless performance of these models when they are used to investigate the evolution of congestion over network during a given time period. To construct a corresponding travel time function for the congested links and then use the obtained function to formulate new TAMs, we can avoid the unreal high flows on the congested links in the situation where the specified flow states of all links are given. To solve the new formulated user equilibrium model which is non-convex, a branch-and-band algorithm is proposed using the partial linearizing technique to deal with the non-convex objective function. To investigate the evolution of congestion over network, we can repeatedly resolve the new formulated model with gradually updated flow states of links. Numerical examples demonstrate the effectiveness of our new approach. By adopting this approach, the traffic assignment result will be more reliable. The other decisions based on the assignment result will be more acceptable and useful.

Many valuable directions can be explored using this new approach as an analysis tool in the future. Based on the investigation of the evolution of congestion over the traffic network, more effective signal control schemes at intersections may be determined during the given time period adapting to the expanding and/or dispersing congested zones. The route guidance system can be more reliable by making use of the information about the reliable link flows and the evolution of congestion over the network. With respect to the issues following the transportation planning, many decisions can be made better. For example, the location of a new infrastructure can be chosen more wisely and the impact of this infrastructure on the surrounding traffic can be estimated more accurately.

## Acknowledgment

This research was supported in part by National Natural Science Foundation of China(71601118, 71801153, 71871144）, the Natural Science Foundation of Shanghai(18ZR1426200), the Shanghai First-class Academic Discipline Project (No. S1201YLXK) and the Key Climbing Project of

USST (No. SK17PA02).